\documentclass{ifacconf}
\usepackage[dcucite]{harvard}
\usepackage[final]{graphics}
\usepackage{amssymb}
\usepackage{epsfig}
\usepackage{amssymb,amsbsy,amsmath,amsfonts,amssymb,amscd}
\usepackage{subfigure}
\usepackage{hhline}
\usepackage[LGR,T1]{fontenc}
\usepackage[greek,english]{babel}
\usepackage{color}

\newtheorem{theorem}{Theorem}[section]
\newtheorem{corollary}[theorem]{Corollary}

\newtheorem{proposition}[theorem]{Proposition}
\newtheorem{remark}[theorem]{Remark}

\begin{document}
\runauthor{Fliess and Join}
\begin{frontmatter}
\title{Time series technical analysis via new fast estimation methods:
a preliminary study in mathematical finance}
\author[Baiae,Paestum]{Michel FLIESS}
\author[Baiae,Rome]{\quad C\'{e}dric JOIN}

\address[Baiae]{INRIA-ALIEN}
\address[Paestum]{LIX (CNRS, UMR 7161),
\'Ecole polytechnique \\ 91128 Palaiseau, France \\ {
Michel.Fliess@polytechnique.edu}}
\address[Rome]{CRAN (CNRS, UMR 7039), Nancy-Universit\'{e} \\
BP 239, 54506 Vand\oe{}uvre-l\`es-Nancy, France
\\ {
Cedric.Join@cran.uhp-nancy.fr} }
\begin{abstract}:
New fast estimation methods stemming from control theory lead to a
fresh look at time series, which bears some resemblance to
``technical analysis''. The results are applied to a typical object
of financial engineering, namely the forecast of foreign exchange
rates, via a ``model-free'' setting, {\it i.e.}, via repeated
identifications of low order linear difference equations on sliding
short time windows. Several convincing computer simulations,
including the prediction of the position and of the volatility with
respect to the forecasted trendline, are provided.
$\mathcal{Z}$-transform and differential algebra are the main
mathematical tools.
\end{abstract}

\begin{keyword}
Time series, identification, estimation, trends, noises, model-free
forecasting, mathematical finance, technical analysis,
heteroscedasticity, volatility, foreign exchange rates, linear
difference equations, $\mathcal{Z}$-transform, algebra.
\end{keyword}
\end{frontmatter}

\section{Introduction}
\subsection{Motivations}
Recent advances in estimation and identification (see, {\it e.g.},
\cite{esaim,cras04,diag,garnier,nl} and the references therein)
stemming from mathematical control theory may be summarized by the
two following facts:
\begin{itemize}
\item Their algebraic nature permits to derive exact non-asymptotic formulae
for obtaining the unknown quantities in real time.
\item There is no need to know the statistical properties of the
corrupting noises.
\end{itemize}
Those techniques have already been applied in many concrete
situations, including signal processing (see the references in
\cite{snr}). Their recent and successful extension to discrete-time
linear control systems \cite{austria} has prompted us to study their
relevance to financial time series.

\begin{remark}
The relationship between time series analysis and control theory is
well documented (see, {\it e.g.}, \cite{box,gou,hamilton} and the
references therein). Our viewpoint seems nevertheless to be quite
new when compared to the existing literature.
\end{remark}

\begin{remark}
The title of this communication is due to its obvious connection
with some aspects of {\em technical analysis}, or {\em charting}
(see, {\it e.g.}, \cite{aronson,bechu,kaufman,kirk,murphy} and the
references therein), which is widely used among traders and
financial professionals.\footnote{Technical analysis is often
severely criticized in the academic world and among the
practitioners of mathematical finance (see, {\it e.g.},
\cite{paulos}).}
\end{remark}

\subsection{Linear difference equations}
Consider the univariate time series $\{x(t) \mid t \in
\mathbb{N}\}$: $x(t)$ is not regarded here as a stochastic process
like in the familiar ARMA and ARIMA models but is supposed to
satisfy ``approximatively'' a linear difference equation
\begin{equation}\label{approx}
x(t + n) - a_1 x(t + n - 1)- \dots - a_n x(t) = 0
\end{equation}
where $a_1, \dots, a_n \in \mathbb{R}$. Introduce as in digital
signal processing the additive decomposition
\begin{equation}\label{nu}
x(t) = x_{\text{trendline}}(t) + \nu(t)
\end{equation}
where
\begin{itemize}
\item $x_{\text{trendline}}(t)$ is the {\em trendline}\footnote{Compare, {\it e.g.},
with \cite{econom}.} which satisfies Eq. \eqref{approx} exactly;
\item the additive ``noise'' $\nu(t)$ is the mismatch between the real data and
the trendline.
\end{itemize}
Thus
\begin{equation}\label{exact}
x(t + n) - a_1 x(t + n - 1) - \dots - a_n x(t) = \epsilon (t)
\end{equation}
where
\begin{equation}\label{noise}
\epsilon (t) = \nu(t + n) - a_1
\nu(t + n-1) - \dots - a_n \nu(t)
\end{equation}
We only assume that the ``ergodic mean'' of $\nu (t)$ is $0$, {\it
i.e.},
\begin{equation}\label{stat}
\lim_{N \to + \infty} \frac{\nu (0) + \nu (1) + \dots + \nu (N)}{N +
1} = 0
\end{equation}
It means that, $\forall ~ t \in \mathbb{N}$, the moving average
\begin{equation}\label{mo}
\text{MA}_{\nu, N} (t) = \frac{\nu (t) + \nu (t + 1) + \dots + \nu
(t + N)}{N + 1}
\end{equation} is close to $0$ if $N$ is large enough. It follows from Eq.
\eqref{noise} that $\epsilon(t)$ also satisfies the properties
\eqref{stat} and \eqref{mo}. Most of the stochastic processes, like
finite linear combinations of i.i.d. zero-mean processes, which are
associated to time series modeling, do satisfy almost surely such a
weak assumption. Our analysis
\begin{itemize}
\item does not make any difference between non-stationary
and stationary time series,
\item does not need the often tedious and cumbersome trend and seasonality
decomposition (our trendlines include the seasonalities, if they
exist).
\end{itemize}

\subsection{A model-free setting}
It should be clear that
\begin{itemize}
\item a concrete time series cannot be ``well'' approximated in general
by a solution of a ``parsimonious'' Eq. \eqref{approx}, {\it i.e.},
a linear difference equation of low order;
\item the use of large order linear difference equations, or of
nonlinear ones, might lead to a formidable computational burden for
their identifications without any clear-cut forecasting benefit.
\end{itemize}
We adopt therefore the quite promising viewpoint of \cite{sm} where
the control of ``complex'' systems is achieved without trying a
global identification but thanks to elementary models which are only
valid during a short time interval and are continuously
updated.\footnote{See the numerous examples and the references in
\cite{sm} for concrete illustrations.} We utilize here low order
difference equations.\footnote{Compare with \cite{willems}.} Then
the window size for the moving average \eqref{mo} does not need to
be ``large''.


\subsection{Content}
Sect. \ref{parameter}, which considers the identifiability of
unknown parameters, extends to the discrete-time case a result in
\cite{snr}. The convincing computer simulations in Sect.
\ref{change} are based on the exchange rates between US Dollars and
\textgreek{\euro}uros. Besides forecasting the trendline, we predict
\begin{itemize}
\item the position of the future rate w.r.t. the forecasted
trendline,
\item the standard deviation w.r.t. the forecasted
trendline.
\end{itemize}
Those results might lead to a new understanding of volatility and
risk management.\footnote{See \cite{taleb} for a critical appraisal
of the existing literature on this subject, which is of utmost
importance in financial engineering. (Extreme) risks are discussed
in \cite{bouchaud,frequency,mandelbrot,sornette} from quite
different perspectives. It is the trendline which would exhibit
abrupt changes in our setting (compare with the probabilistic
standpoint; see, {\it e.g.}, \cite{wilmott} and the references
therein). Our estimation techniques permit an efficient change-point
detection \cite{ajaccio}, which needs to be extended, if possible,
to some kind of forecasting.} Sect. \ref{conclusion} concludes with
a short discussion on the notion of {\em trend}.

\section{Parameter identification}\label{parameter}

\subsection{Rational generating functions}\label{rationel}
Consider again Eq. \eqref{approx}. The $\mathcal{Z}$-transform $X$
of $x$ satisfies (see, {\it e.g.}, \cite{doetsch,jury}) {\small
\begin{equation}\label{operation}
\begin{array}{l}
z^n [X - x(0) - x(1)z^{-1} - \dots - x(n - 1)z^{-(n - 1)}]  \\
- \dots - a_{n-1} z [X - x(0)] - a_n X = 0
\end{array}
\end{equation}}
It shows that $X$, which is called the {\em generating function} of
$x$, is a rational function of $z$, {\it i.e.}, $X \in \mathbb{R}
(z)$:
\begin{equation}\label{rational}
X = \frac{P(z)}{Q(z)}
\end{equation}
where
$$
\begin{array}{l}
P(z) = b_0 z^{n-1} + b_1 z^{n-2} + \dots + b_{n-1} \in \mathbb{R}[z]
\\ Q(z) = z^n - \dots - a_{n-1}z - a_n  \in \mathbb{R}[z]
\end{array}
$$
Hence
\begin{proposition}\label{rationnel}
$x(t)$, $t \geq 0$, satisfies a linear difference equation
\eqref{approx} if, and only if, its generating function $X$ is a
rational function.
\end{proposition}
It is obvious that the knowledge of $P$ and $Q$ permits to determine
the initial conditions $x(0), \dots, x(n-1)$.

\begin{remark}
Consider the inhomogeneous linear difference equation
$$
\begin{array}{l}
x(t + n) - a_1 x(t + n - 1)- \dots - a_n x(t)
\\ = \sum_{\text{finite}} \varpi (t) \alpha^t +
\sum_{\text{finite}} \varpi^\prime (t) \sin (\omega t + \varphi)
\end{array}
$$
where $\varpi (t), \varpi^\prime (t) \in \mathbb{R}[t]$, $\alpha,
\omega, \varphi \in \mathbb{R}$. Then the $\mathcal{Z}$-transform $X
\in \mathbb{R} (z)$ of $x(t)$ is again rational. It is equivalent
saying that $x(t)$, $t\geq 0$, still satisfies a homogeneous
difference equation.
\end{remark}


\subsection{Parameter identifiability}
\subsubsection{Generalities}
Let $$\mathfrak{K} = \mathbb{Q}\left(a_1, \dots, a_n,
b_0, \dots, b_{n-1}\right) $$ be the field generated over the field
$\mathbb{Q}$ of rational numbers by $a_1, \dots, a_n, b_0, \dots,
b_{n-1}$, which are considered as unknown parameters and therefore
in our algebraic setting as independent indeterminates
\cite{esaim,diag,garnier}. Write $\bar{\mathfrak{K}}$ the algebraic
closure of $\mathfrak{K}$ (see, {\it e.g.}, \cite{lang,chambert}).
Then $X \in \bar{\mathfrak{K}}(z)$, {\it i.e.}, $X$ is a rational
function over $\bar{\mathfrak{K}}$. Moreover $\bar{\mathfrak{K}}(z)$
is a {\em differential field} (see, {\it e.g.}, \cite{chambert})
with respect to the derivation $ \frac{d}{dz} $. Its subfield of
{\em constants} is the algebraically closed field
$\bar{\mathfrak{K}}$.

Introduce the square Wronskian matrix $\mathcal{M}$ of order $2n +
1$ \cite{chambert} where its $\chi^{th}$-row, $0 \leq \chi \leq 2n$,
is
\begin{equation}\label{row}
\frac{d^\chi}{dz^\chi} z^n X, \dots, \frac{d^\chi}{dz^\chi}  X,
\frac{d^\chi}{dz^\chi} z^{n - 1}, \dots, \frac{d^\chi}{dz^\chi} 1
\end{equation}
It follows from Eq. \eqref{operation} that the rank of $\mathcal{M}$
is $2n$ if, and only if, $x$ does not satisfy a linear difference
relation of order strictly less than $n$. Hence
\begin{theorem} \label{linident}
If $x$ does not satisfy a linear difference equation of order
strictly less than $n$, then the parameters $$a_1, \dots, a_n, b_0,
\dots, b_{n-1}$$ are {\em linearly identifiable}.\footnote{It means
following the terminology of \cite{esaim,garnier} that $a_1, \dots,
a_n, b_0, \dots, b_{n-1}$ are uniquely determined by a system of
$2n$ linear equations, the coefficients of which depend on
$\frac{d^\chi}{dz^\chi} X$ and $\frac{d^\chi}{dz^\chi} z^m$, $0 \leq
m \leq n-1$.}
\end{theorem}


\subsubsection{Identifiability of the dynamics} For identifying the
dynamics, {\it i.e.}, $a_1, \dots, a_n$, without having to determine
the initial conditions consider the $(n+1) \times (n+1)$ Wronskian
matrix $\mathcal{N}$, where its $\mu^{th}$-row, $0 \leq \mu \leq
n+1$, is
$$ \frac{d^{n + \mu}}{dz^{n + \mu}} z^n X, \dots, \frac{d^{n + \mu}}{dz^{n + \mu}}  X $$
It is obtained by taking the $X$-dependent entries in the $n+1$ last
rows of type \eqref{row}, {\it i.e.}, in disregarding the entries
depending on $b_0, \dots, b_{n-1}$. The rank of $\mathcal{N}$ is
again $n$. Hence
\begin{corollary}
$a_1, \dots, a_n$ are linearly identifiable.
\end{corollary}

\subsubsection{Identifiability of the numerator} Assume now that the
dynamics is known but not the numerator $P$ in Eq. \eqref{rational}.
We obtain $b_0, \dots, b_{n-1}$ from the first $n$ rows \eqref{row}.
Hence
\begin{corollary}
$b_0, \dots, b_{n-1}$ are linearly identifiable.
\end{corollary}

\subsection{Some hints on the computer implementation}\label{hint}
We proceed as in \cite{esaim,garnier} and in \cite{austria}. The
unknown linearly identifiable parameters are solutions of a matrix
linear equation, the coefficients of which depend on $x$. Let us
emphasize that we substitute to $x$ its filtered value thanks to a
discrete-time version of \cite{MED07}.\footnote{See, {\it e.g.},
\cite{gen} for an excellent presentation of various filtering
techniques in economics and finance.}

\section{Example: Forecasting 5 days ahead the {\$} - \textgreek{\euro} exchange
rates}\label{change} 
\begin{figure}[!ht]
\centering
{{\rotatebox{-90}{\resizebox{!}{9.5cm}{%
\includegraphics{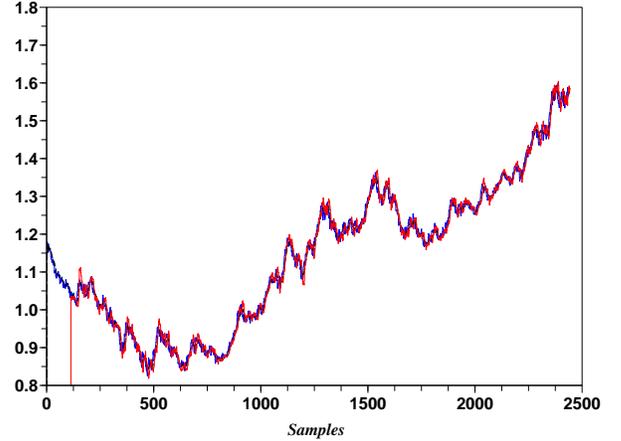}}}}}
\caption{Exchange rates (blue --), filtered signal (black - -),
forecasted signal ($5$ days ahead) (red --) \label{US5a}}
\end{figure}

\begin{figure*}[!ht]
\centering {{\rotatebox{-90}{\resizebox{!}{13.4cm}{
\includegraphics{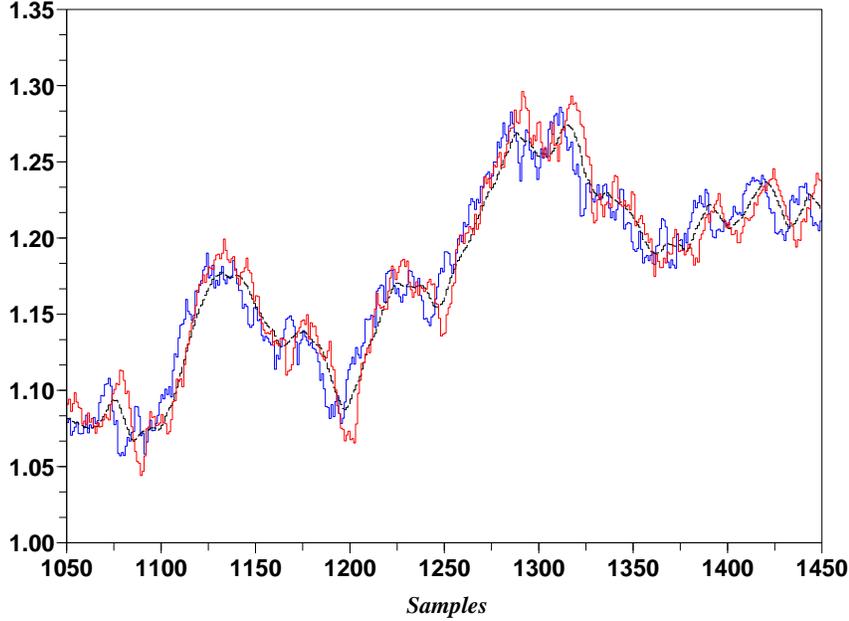}}}}}
\caption{Zoom of Figure \ref{US5a} \label{US5b}}
\end{figure*}

We are utilizing data from the European Central
Bank, depicted by the blue lines in the Figures \ref{US5a} and
\ref{US5b}, which summarize the $2400$ last daily exchange rates
between the US Dollars and the \textgreek{\euro}uros.\footnote{The
authors are perfectly aware that only computations dealing with high
frequency data might be of practical value. This type of results
will be presented elsewhere.}

\subsection{Forecasting the trendline}\label{fotr}
\begin{figure}[!ht] \centering
{{\rotatebox{-90}{\resizebox{!}{9.5cm}{%
\includegraphics{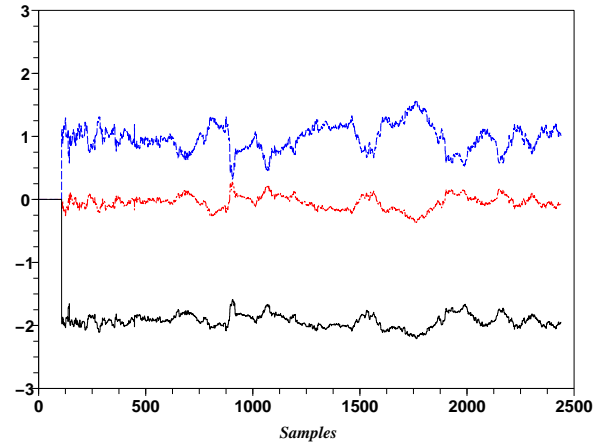}}}}}
\caption{Parameter estimations $a_1$ (red $-.-$), $a_2$ (blue $- -$)
and $a_3$ (black $--$) ($5$ days ahead) \label{US5c}}
\end{figure}

In order to forecast the exchange rate $5$ days ahead we apply the
rules sketched in Sect. \ref{hint} and we utilize a linear
difference equation \eqref{approx} of order $3$ (the filtered values
of the exchange rates are given  by the black lines in the Figures
\ref{US5a}, \ref{US5b}). Fig. \ref{US5c} provides the estimated
values of the coefficients of the difference equation. The results
on the forecasted values of the exchange rates are depicted by the
red lines in the Figures \ref{US5a} and \ref{US5b}, which should be
viewed as a predicted trendline.

\subsection{Above or under the predicted trendline?}\label{above}
\begin{figure}[!ht]
\centering
{{\rotatebox{-90}{\resizebox{!}{9.5cm}{%
\includegraphics{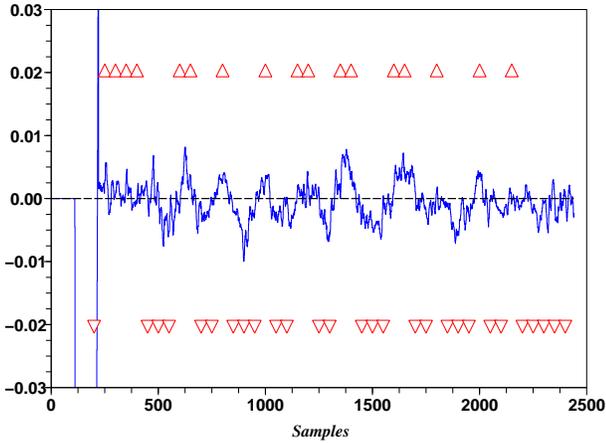}}}}}
\caption{Predicted position w.r.t. the trendline ($5$ days ahead) --
$\nabla$: above, $\triangle$: under \label{US5d}}
\end{figure}

Consider again the ``error'' $\nu (t)$ in Eq. \eqref{nu} and its
moving average $\text{MA}_{\nu, N} (t)$ in Eq. \eqref{mo}.
Forecasting $\text{MA}_{\nu, N} (t)$ as in Sect. \ref{fotr} tells us
an expected position with respect to the forecasted trendline. The
blue line of Fig. \ref{US5d} displays the result for the window size
$N = 100$. The meaning of the {\em indicators} $\triangle$ and
$\nabla$ is clear.

Table \ref{tUSe} compares for various window sizes the signs of the
predicted values of $\text{MA}_{\nu, N} (t)$, which tells us if one
should expect to be above or under the trendline, with the true
positions of $x(t)$ with respect to the trendline. The results are
expressed via percentages.

\begin{table}
\center\begin{tabular}{|c|c|c|} \hline Window's size & Percentage
\\ \hhline{==} $50$ & $65.6\%$
\\\hline $100$ &$88.3\%$ \\\hline $200$ & $62.3\%$ \\\hline $300$ &
$67.1\%$ \\\hline
\end{tabular}
{\tiny \caption{Comparison between the sign of the predicted value
of $\text{MA}_{\nu, N} (t)$ and the true position of $x(t)$ w.r.t.
the trendline ($5$ days ahead).} \label{tUSe}}
\end{table}

\subsection{Predicted volatility w.r.t. the
trendline}\label{vola} 
\begin{figure}[!ht]
\centering
{{\rotatebox{-90}{\resizebox{!}{9.5cm}{%
\includegraphics{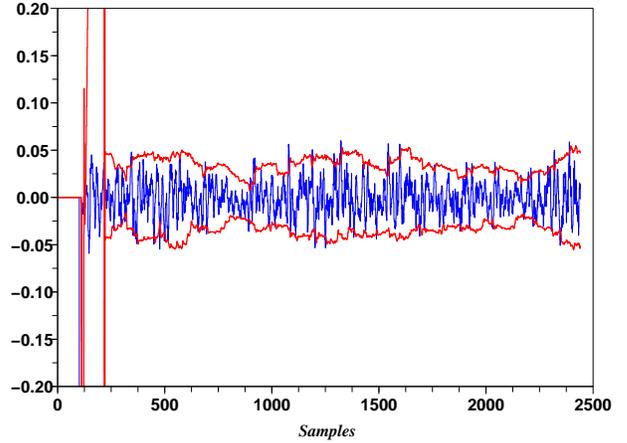}}}}}
\caption{ Confidence interval ($95 \%$) ($5$ days ahead)
\label{US5e}}
\end{figure}

Introduce the {\em moving standard deviation}
\begin{equation*}\label{std}
{\small \begin{array}{l}
\text{MSTD}_{\nu,N}(t) = \\
\sqrt{\frac{\sum_{\tau = 0}^{N}(\nu (t + \tau) - \text{MA}_{\nu, N}
(t - N + \tau))^2}{N+1}}
\end{array}}
\end{equation*}
and forecast it as in Sect. \ref{above}. The results, which are
displayed for a window size $N=100$ in  Table 2 and Fig. \ref{US5e}
via the familiar confidence intervals,\footnote{There is of course
no need for the underlying statistics to be Gaussian. Lack of space
prevents us from exhibiting forecasts of quantities like {\em
skewness} and {\em kurtosis}, which would be obtained by similar
calculations. This will be done in some future publications.}
confirm the time-dependence of the variance, {\it i.e.}, the {\em
heteroscedasticity}.
\begin{table}
\center\begin{tabular}{|c|c|c|} \hline
Confidence intervals& Prediction & Real\\
 \hhline{===}
\text{mean-$3 \times$std,mean+$3\times$std}&$99\%$ & $98.7\%$
\\\hline \text{mean-$2\times$std,mean+$2\times$std}&$95\%$ &$92.2\%$ \\\hline
\text{mean-std,mean+std}&$68\%$ & $64.4\%$ \\\hline
\end{tabular}
\caption{Confidence interval validations ($5$ days ahead)
\label{tUSs}}
\end{table}

\subsection{Forcasting 10 days ahead}
\begin{figure}[!ht]
\centering
{{\rotatebox{-90}{\resizebox{!}{9.5cm}{%
\includegraphics{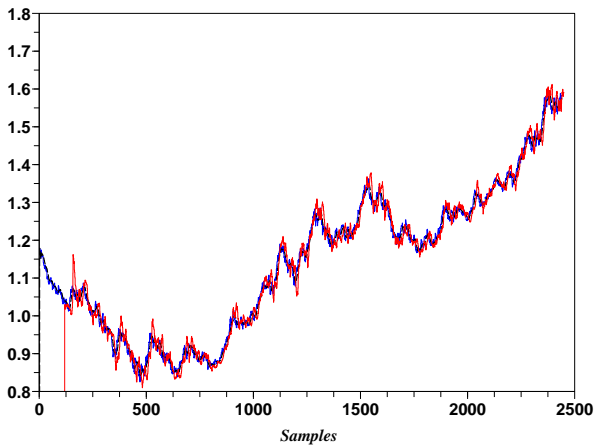}}}}}
\caption{Exchange rates (blue --), filtered signal (black - -),
forecasted signal (red --)  ($10$ days ahead) \label{US10a}}
\end{figure}

\begin{figure*}[!ht]
\centering
{{\rotatebox{-90}{\resizebox{!}{13.4cm}{%
\includegraphics{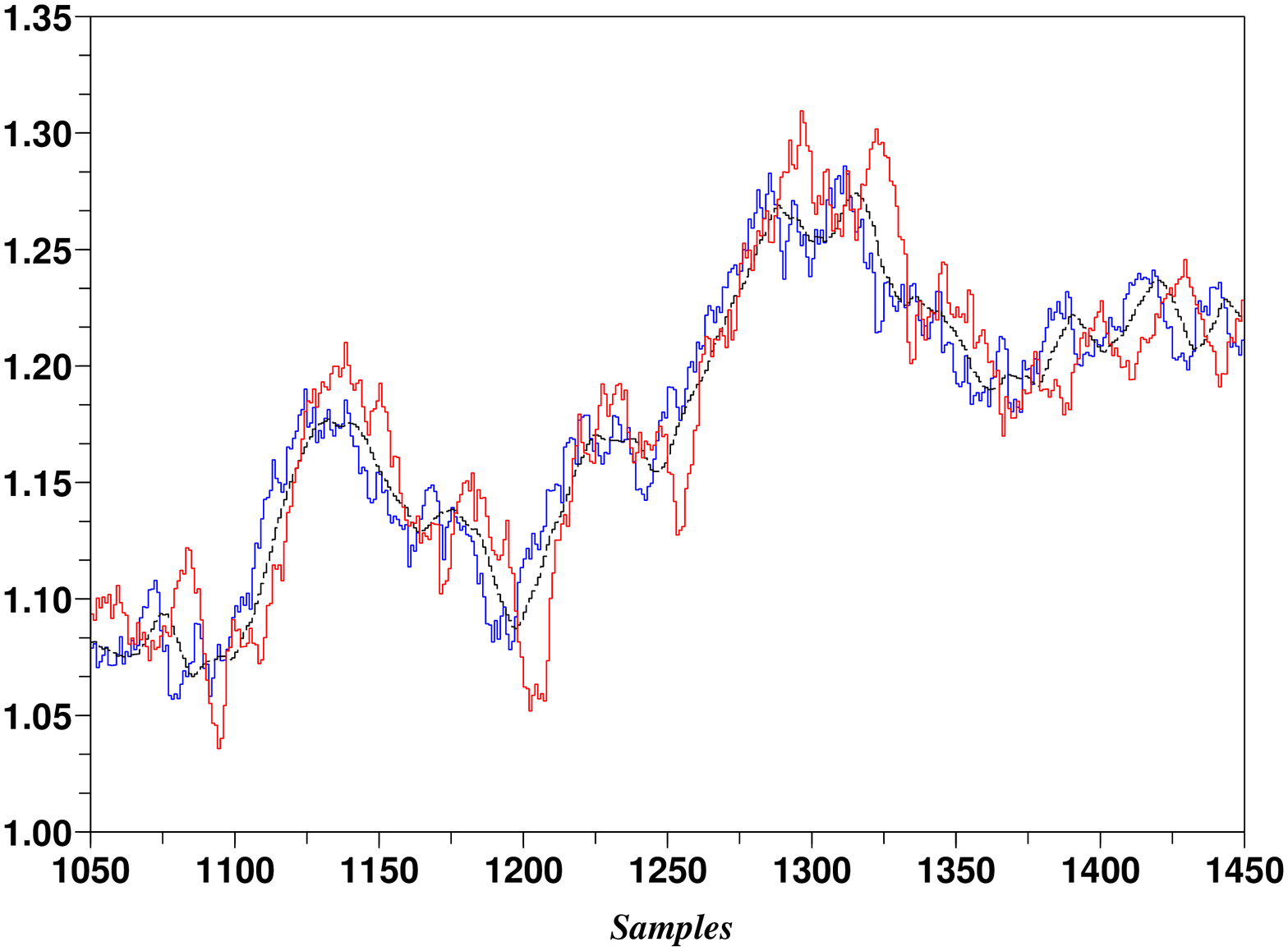}}}}}
\caption{Zoom of Figure \ref{US10a} \label{US10b}}
\end{figure*}

\begin{figure}[!ht]
\centering
{{\rotatebox{-90}{\resizebox{!}{9.5cm}{%
\includegraphics{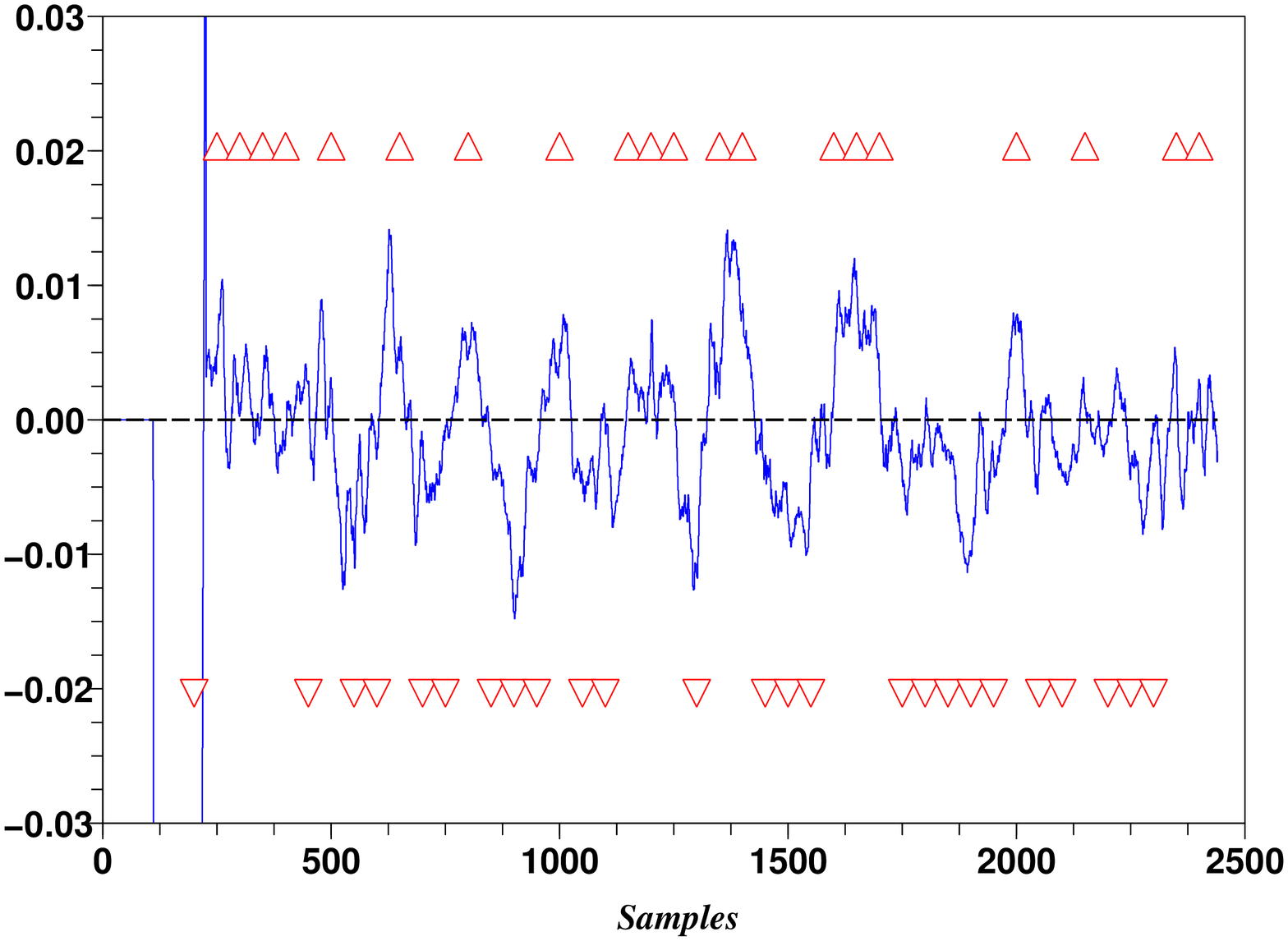}}}}}
\caption{Predicted position w.r.t. the trendline ($10$ days ahead)
-- $\nabla$: above, $\triangle$: under \label{US10d}}
\end{figure}

\begin{figure}[!ht]
{{\rotatebox{-90}{\resizebox{!}{9.5cm}{%
\includegraphics{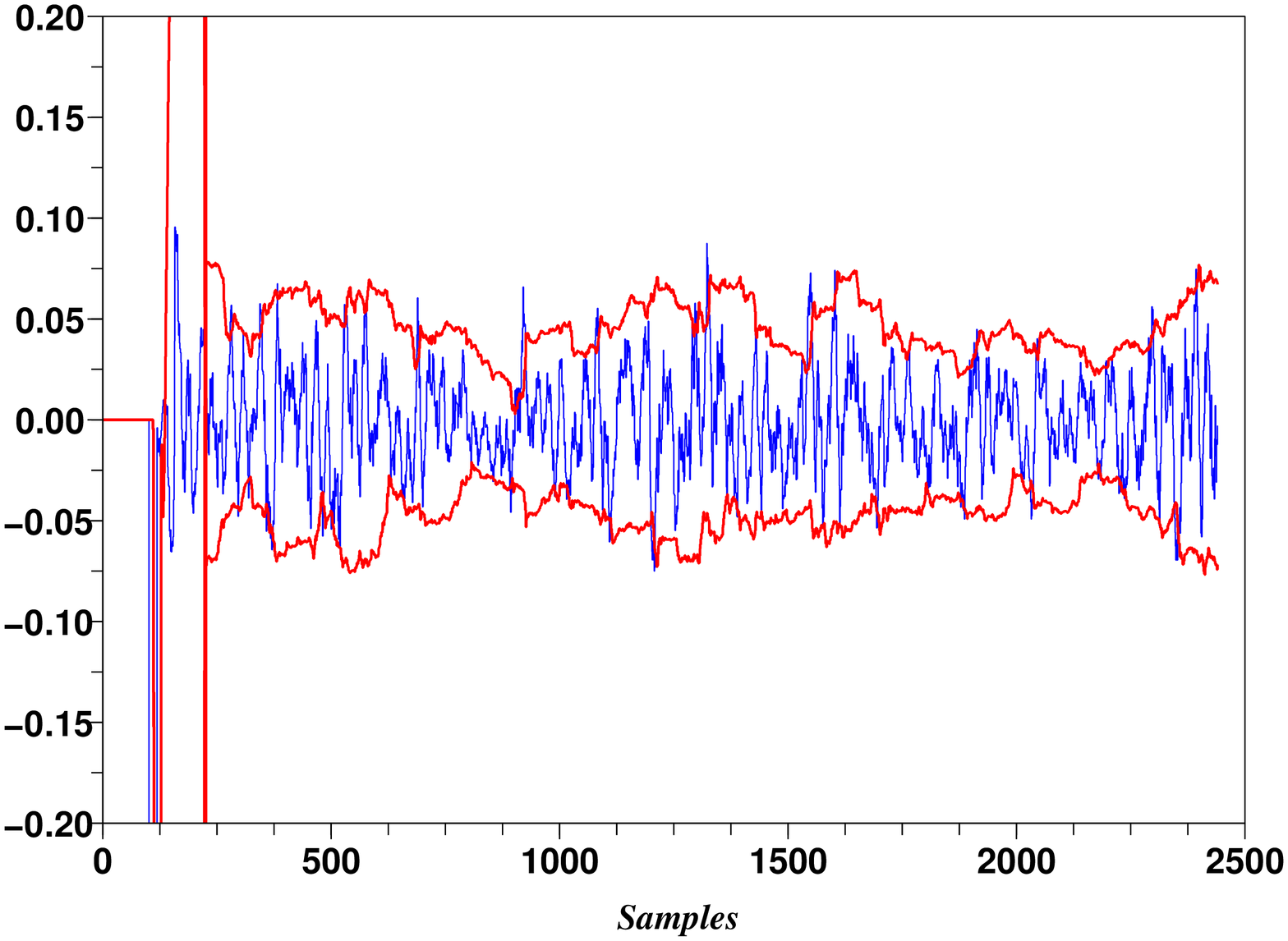}}}}}
\caption{ Confidence interval ($95 \%$) ($10$ days ahead)
\label{US10e}}
\end{figure}

Figures \ref{US10a}, \ref{US10b}, \ref{US10d}, \ref{US10e} display
the same type of results as in Sections \ref{fotr}, \ref{above},
\ref{vola} via similar computations for a forecasting $10$ days
ahead. The quality of the computer simulations only slightly
deteriorates.

\section{Conclusion}\label{conclusion}
The existence of {\em trends}, which is
\begin{itemize}
\item the key assumption in technical analysis,\footnote{Trends in technical
analysis should not be confused with what are called {\em trends} in
the time series literature (see, {\it e.g.}, \cite{gou,hamilton}).}
\item quite foreign, to the best of our knowledge,
to the academic mathematical finance, where the paradigm of {\em
random walks} is prevalent (see, {\it e.g.}, \cite{wilmott}),
\end{itemize}
is fundamental in our approach. A theoretical justification will
appear soon \cite{trend}.\footnote{The existence of trends does not
necessarily contradict a random character (see \cite{trend} for
details).} We hope it will lead to a sound foundation of technical
analysis,\footnote{See also \cite{frequency} for a most exciting
study which employs high frequency data. There are also other types
of attempts to put technical analysis on a firm basis (see, {\it
e.g.}, \cite{lo}). See \cite{talay} for a comparison between
technical analysis and model-based approaches with parametric
uncertainties.} which will bring as a byproduct easily implementable
real-time computer programs.\footnote{Our technics already lead to
such computer programs in automatic control and in signal
processing.}

\noindent{\small{\bf Acknowledgement}. The authors wish to thank G.
Daval-Leclercq (Soci\'{e}t\'{e} G\'{e}n\'{e}rale - Corporate \& Investment Banking)
for helpful discussions and comments.}


\end{document}